\documentclass{jfm}
\usepackage{graphicx,ulem}
\usepackage{epstopdf, epsfig}
\usepackage{amsmath}
\usepackage{amsfonts}
\usepackage{amssymb}

\usepackage{xcolor}

\shorttitle{Axisymmetric internal wave transmission and resonance in non-linear stratifications}
\shortauthor{S. Boury, P. Odier and T. Peacock}

\title{Axisymmetric internal wave transmission and resonance in non-linear stratifications}

\author{S. Boury\aff{1}
  \corresp{\email{samuel.boury@ens-lyon.fr}},
  P. Odier\aff{1}
 \and T. Peacock\aff{2}}

\affiliation{\aff{1}Univ Lyon, ENS de Lyon, Univ Claude Bernard, CNRS, Laboratoire de Physique, F-69342 Lyon, France
\aff{2}Department of Mechanical Engineering, Massachusetts Institute of Technology, Cambridge, MA 02139, USA}

\begin{document}

\maketitle

\begin{abstract}
To date, the influence of non-linear stratifications and two layer stratifications on internal wave propagation has been studied for two-dimensional wave fields in a cartesian geometry. Here, we use a novel wave generator configuration to investigate transmission in non-linear stratifications of axisymmetric internal wave. Two configurations are studied, both theoretically and experimentally. In the case of a free incident wave, a transmission maximum is found in the vicinity of evanescent frequencies. In the case of a confined incident wave, resonant effects lead to enhanced transmission rates from an upper layer to layer below. We consider the oceanographic relevance of these results by applying them to an example oceanic stratification, finding that there can be real-world implications.
\end{abstract}

\begin{keywords}
Authors should not enter keywords on the manuscript, as these must be chosen by the author during the online submission process and will then be added during the typesetting process (see http://journals.cambridge.org/data/\linebreak[3]relatedlink/jfm-\linebreak[3]keywords.pdf for the full list)
\end{keywords}

% ################################################################################ %
	\section{Introduction}
	
	Inertia-gravity waves are known to be a significant mechanism for energy and momentum transfer in the ocean and atmosphere~\citep{sutherland2010}. While a substantial amount of progress in understanding of their roles has been achieved via plane wave models, a planar geometry is not necessarily appropriate for studies pertaining to natural phenomena such as oceanic internal wave generation by hurricanes~\citep{kumar2011} and tropical cyclones~\citep{schubert1980}, and atmospheric generation by thunderstorms~\citep{sentman2003}. Furthermore, \cite{warren1960} noted that a vertical disturbance of small spatial extent compared to the horizontal dimension of the system in a stably stratified fluid like the atmosphere leads to the emission of an axisymmetric perturbation. As such, axisymmetry is a reasonable approximation to consider for inertia-gravity waves generated by localized disturbances in either the atmosphere or the ocean.
	
	\cite{mowbray1967} and \cite{appleby1987} studied immersed objects such as spheres as being sources of radiated axisymmetric internal waves. Several subsequent laboratory experimental studies  investigated axisymmetric inertia-gravity waves,  focusing primarily on the global structure of the cone-shaped internal wave emission induced by an oscillating or rotating object~\citep{peacock2005, duranmatute2013, ghaemsaidi2013, dizes2015}. More recently, ~\cite{ansong2010} investigated internal waves excited by an axisymmetric convective plume in a constant density-gradient stratification. To obtain a high degree of control on the spatial form of an axisymmetric source, \cite{maurer2017} adapted the oscillating plate planar generator of ~\cite{gostiaux2006} to produce a wave generator capable of producing axisymmetric internal wave fields of arbitrary radial form, such as truncated Bessel modes or annular patterns, which opens the door on a wide range of possible experimental investigations. Inspired by studies on confined two dimensional internal waves, \cite{boury2018}~used this apparatus to produce pure axisymmetric modes that can be used as the basis for understanding the behaviour of more complex fields via modal decomposition using the natural cylindrical basis of Bessel functions. 
	
	 In the oceans and atmosphere an important consideration is the vertical form of the density-gradient. The influence of non-uniform stratifications has been studied for planar geometry, determining transmission and reflection coefficients for stratifications with sharp~\citep{nault2007,mathur2009} and smooth ~\citep{brown2007, mathur2009} discontinuities. Numerical and experimental studies show good agreement with the theoretical predictions, considering two different scenarios: one with freely propagating incident waves~\citep{sutherland2004, mathur2009}, a configuration relevant to the atmosphere, and another with confined wave resonance in a forced, upper stratification layer~\citep{ghaemsaidi2016}, relevant to an ocean configuration. 
	 
	 To date, all experimental studies of axisymmetric internal wave fields have considered linearly stratified fluids and there have been no such transmission studies for axisymmetric inertia-gravity wave fields. In this paper, therefore, we delve further into the propagation properties of axisymmetric internal waves in non-linear stratifications. The plane wave analysis of ~\cite{mathur2009} is extended to cylindrical wave modes and we derive a theoretical prediction for their transmission across a buoyancy frequency interface. Using the experimental configuration of \cite{boury2018}, experimental transmission studies are conducted for a freely propagating axisymmetric internal wave and for an axisymmetric internal wave field excited in an upper layer; in both cases the radial form of the wave field is a mode $1$ Bessel function. In section $2$, we derive the theoretical framework and define the boundary value problem used for numerical computation. Then, in section $3$, we describe the experimental apparatus. Our experimental results are presented in section $4$, followed by oceanographic considerations in section $5$ and conclusions and discussion in section $6$.
	
% ################################################################################ %
	\section{Theory}
		\subsection{Governing Equations}
		
			Focusing on an inviscid Boussinesq fluid rotating at an angular velocity $\Omega$, with a density stratification $\rho(z)$ where $z$ is the ascendent vertical, small amplitude internal waves satisfy the following equation in cylindrical coordinates~\citep{boury2018, maurer2017}
			\begin{equation}
				\dfrac{\partial^2}{\partial t^2}\left( \dfrac{\partial^2 \psi}{\partial z^2} + \dfrac{\partial}{\partial r}\left( \frac{1}{r} \dfrac{\partial (r \psi)}{\partial r} \right) \right) = -f^2 \dfrac{\partial^2 \psi}{\partial z^2} - N^2 \dfrac{\partial}{\partial r}\left( \frac{1}{r} \dfrac{\partial (r \psi)}{\partial r} \right),
				\label{eq:IW01}
			\end{equation}
			in which $f = 2\Omega$ is the Coriolis frequency, and $N = \sqrt{(-g/\rho_0) \partial \bar{\rho} / \partial z}$ is the buoyancy frequency, $\rho_0$ being the reference density. The stream function $\psi$ is defined so that the radial and vertical velocities $v_r$ and $v_z$ are
			\begin{equation}
				v_r = - \frac{1}{r}\dfrac{\partial (r \psi)}{\partial z} \mathrm{~~~~~~~and~~~~~~~} v_z =\frac{1}{r}\dfrac{\partial (r \psi)}{\partial r}.
				\label{eq:IW02}
			\end{equation}
	
			In the configuration of a horizontal forcing in a linear stratification ($N$ is a constant), the solutions of equation~\eqref{eq:IW01} can be expressed as a product of two decoupled functions: a radial part depending on the radius $r$, and a vertical part depending on the depth $z$. The cylindrical geometry leads to a natural decomposition over Bessel functions for the radial component and exponentials for the vertical component. Using a complex notation, the stream function can therefore be written as a sum of cylindrical modes of the form \citep{boury2018}
			\begin{equation}
				\psi (r,z,t) = \psi_0 J_1(lr) e^{i(mz - \omega t)},
				\label{eq:IW03}
			\end{equation}
			with $\psi_0$ a constant amplitude, $\omega$ the wave frequency, and $l$ and $m$ its radial and vertical wavenumbers. The radial dependence is expressed through a first order Bessel function of the first kind $J_1$, corresponding to a non-divergent mode. In the case of either a free propagating or evanescent wave, the vertical dependence can be written with an exponential. For a more general framework in further sections, for example while considering vertically confined wave fields, the vertical component will be noted $\phi$.  As in cartesian geometry, the parameters $\omega$, $l$, and $m$ are linked through the internal wave dispersion relation \citep{boury2018, maurer2017}
			\begin{equation}
				m = \pm l \left(\frac{\omega^2 - N^2}{f^2 - \omega^2}\right)^{1/2},
				\label{eq:IW04}
			\end{equation}
			derived from~\eqref{eq:IW01},~\eqref{eq:IW02} and \eqref{eq:IW03}.
			
\begin{figure}
				\centering
				\epsfig{file=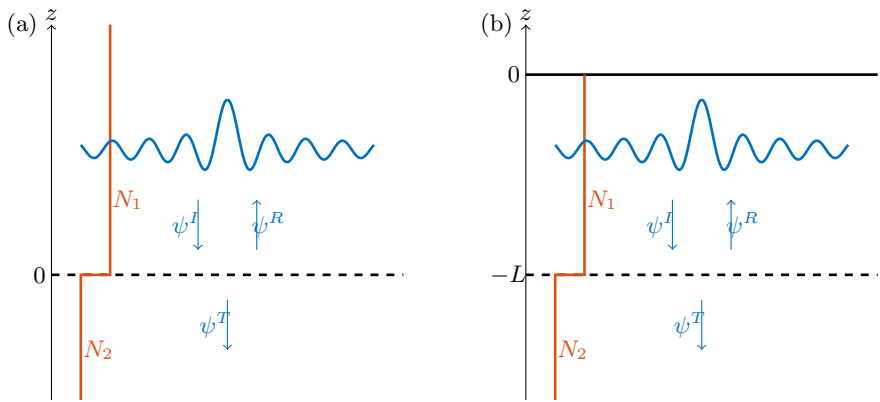} 
				\caption{Transmission of a radial mode $J_1$ across a buoyancy frequency interface. In both cases, the interface is modeled by a sharp discontinuity in buoyancy frequency. (a) free incident wave $\psi^I$ reaching the interface, leading to a reflected wave $\psi^R$ and a transmitted wave $\psi^T$. (b) a confined incident wave $\psi^I$, located between the surface at $z=0$ and the interface at $z=-L$, is an infinite sum of waves being reflected at $z=0$ and at $z=-L$, and leads to a transmitted wave $\psi^T$.}
				\label{fig1}
			\end{figure}

			In the next two sections, we discuss the two scenarios presented in figure~\ref{fig1}, initially assuming a sharp interface (discontinuity between two different buoyancy frequencies $N_1$ and $N_2$, with $N_1>N_2$). Figure~\ref{fig1}(a) shows the transmission of a free incident wave $\psi^I$, which is more relevant to the modeling of atmospheric internal waves, whereas figure~\ref{fig1}(b) shows the transmission of an incident wave confined from above, which is closer to an oceanic situation with a surface forcing.
			
% ------------------------------------------------------------------------------------- %
		\subsection{Transmission of a Free Incident Wave}
			\subsubsection{Sharp Interface}
			
				Linear vertical propagation of radial modes has been extensively studied by \cite{boury2018}. In this section, we consider a horizontal interface located at $z=0$ that splits the domain into two media of constant buoyancy frequency $N$: $N_1$ in the upper layer ($z>0$), $N_2$ in the lower layer ($z<0$), the density being continuous at the interface (figure~\ref{fig1}(a)). We assume a larger buoyancy frequency in the upper layer so that $N_1>N_2$. As depicted in figure~\ref{fig1}(a), a free incident wave $\psi^I$ reaching the inteface leads to a reflected wave $\psi^R$ in the upper region, and a transmitted wave $\psi^T$ in the lower region. The total wavefield can be described by $\psi_1 = \psi^I + \psi^R$ in the upper region, where $\psi^I$ is known, and by $\psi_2 = \psi^T$ in the lower region.
				
				We define the transmission coefficients as the ratios of the velocities above and under the interface, as
				\begin{equation}
					T_{v_z} = \left| \frac{v_z^T}{v_z^I} \right|\mathrm{~~~~~~~and~~~~~~~}     T_{v_r} = \left| \frac{v_r^T}{v_r^I} \right|,
				\end{equation}
				and
				\begin{equation}
					R_{v_z} = \left| \frac{v_z^R}{v_z^I} \right|\mathrm{~~~~~~~and~~~~~~~}     R_{v_r} = \left| \frac{v_r^R}{v_r^I} \right|,
				\end{equation}
				where the notation $I$, $R$, and $T$, indicates that the vertical velocity $v_z$ or the radial velocity $v_r$ is computed from the stream function $\psi^I$, $\psi^R$, and $\psi^T$, respectively (see figure~\ref{fig1}). These velocities are taken at the interface, which means in the limit $z\rightarrow 0$.
				
\begin{figure}
					\centering
					\epsfig{file=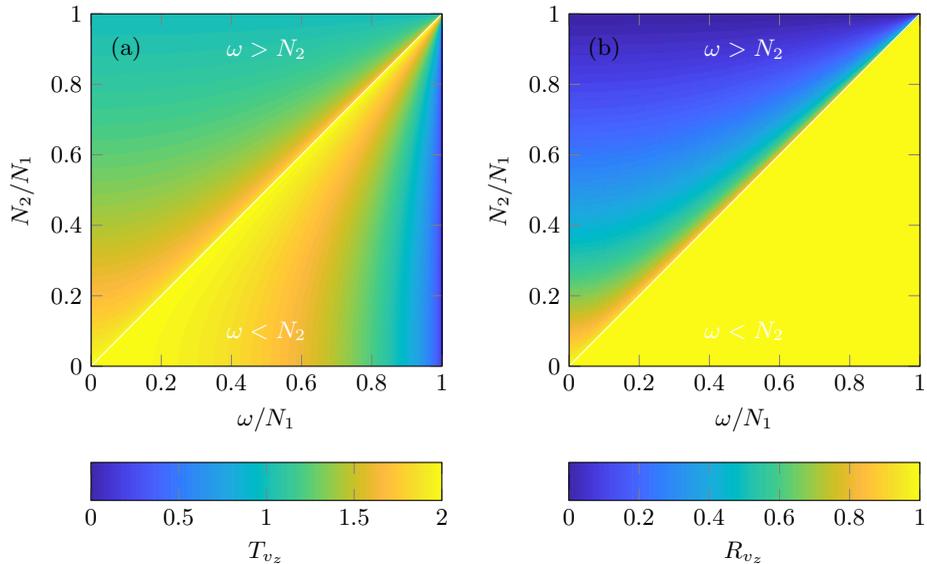} 
					\caption{Colormaps of (a) the transmission coefficient $T_{v_z}$ and (b) the reflection coefficient $R_{v_z}$ in vertical velocity, in the plane $(\omega / N_1, N_1/N_2)$. The Coriolis frequency is set to be $f=0$.}
					\label{fig2}
				\end{figure}

				Transmission and reflection properties can be derived by applying boundary conditions at the interface: continuity of the total vertical velocity field $v_z$ and the pressure $p$ derived from $\psi_1$ and $\psi_2$~\citep{mathur2009}. As in cartesian geometry, these quantities can be obtained using the linearized Navier-Stokes equations~\citep{ansong2010}. The transmission and reflection coefficients for the radial and vertical velocities $v_r$ and $v_z$, and the flux of the energy $j = p v_z$ across the interface, can therefore be derived, respectively
				\begin{equation}
					T_{v_r} = \left\vert \frac{2 m_2}{m_1 + m_2} \right\vert, \mathrm{~~~~~~~} T_{v_z} = \left\vert \frac{2 m_1}{m_1 + m_2}\right\vert, \mathrm{~~~~~~~and~~~~~~~} T_{j} = \left\vert \frac{4 m_1 m_2}{(m_1 + m_2)^2}\right\vert,
				\label{eq:IW05}
				\end{equation}
				and
				\begin{equation}
					R_{v_r} = \left\vert \frac{m_1 - m_2}{m_1 + m_2}\right\vert, \mathrm{~~~~~~~} R_{v_z} = \left\vert \frac{m_2 - m_1}{m_1 + m_2}\right\vert, \mathrm{~~~~~~~and~~~~~~~} R_{j} = \left\vert \frac{(m_1 - m_2)^2}{(m_1 + m_2)^2}\right\vert,
				\label{eq:IW06}
				\end{equation}
				where $m_1$ and $m_2$ are the vertical wavenumbers corresponding to regions $1$ and $2$, given by the dispersion relation~\eqref{eq:IW04}. These coefficients are the same as in cartesian geometry for a plane wave beam crossing a density gradient interface~\citep{mathur2009}, suggesting that the radial dependence does not affect wave properties at the interface though it still shapes the wavefield and the energy distribution. Colormaps of the transmission and reflection coefficient for the vertical velocity, $T_{v_z}$ and $R_{v_z}$, are presented in figure~\ref{fig2} as functions of $N_1/N_2$ and $\omega/N_1$. A maximum of transmission of vertical velocity appears for $\omega/N_1=N_2/N_1$, when the wave just becomes evanescent in the lower region. This evanescence transition is clearly identified in the reflection coefficient, showing a total reflection of the wave field for $\omega/N_1 > N_2/N_1$.
		
% ..................................................................................... %
			\subsubsection{Smooth Interface}\label{smooth}
			
				Following the approach of~\cite{mathur2009}, the propagation of the wave across a physical smooth interface of finite-width transition region $\delta$ is investigated using the buoyancy frequency profile
				\begin{equation}
					N^2(z) = \left(\frac{N_1^2 + N_2^2}{2} \right) + \left(\frac{N_1^2 - N_2^2}{2}\right) \tanh \left(\frac{z}{\delta}\right),
					\label{eq:IW07}
				\end{equation}
				for which $N$ is assumed to be constant far from the transition region and continuously going from $N_1$ to $N_2$ at the interface located at $z=0$.
		
				In a similar way as done by~\cite{mathur2009} and~\cite{nault2007}, the wave behaviour close to the interface is explored through a numerical approach. As the buoyancy frequency profile does only depend on $z$, the radial part is still described by a first order Bessel function $J_1$. With the use of an ansatz describing the stream function as $\psi (r,z,t) = \phi (z) J_1 (l r) e ^{i\omega t}$, the vertical dependence $\phi$ is found to satisfy the differential equation
				\begin{equation}
					\phi '' + \gamma (z) \phi = 0 \mathrm{~~~~~~~with~~~~~~~} \gamma (z) = l^2 \left(\frac{\omega^2 - N^2 (z)}{f^2 - \omega^2}\right),
					\label{eq:IW08}
				\end{equation}
				where $\phi''$ stands for the second order derivative of $\phi$, and $\gamma$ is a function of $z$ that replaces the $m$ wave number given in the sharp interface study. Far from the interface, the wave field is expected to behave vertically as a Fourier mode, which implies an asymptotic form
				\begin{equation}
					\tilde{\phi}(z) = 
						\begin{cases}
							\phi_0 ^I e^{i m_1 z} + \phi_0 ^R e^{-i m_1 z}	& \mathrm{if~~} z = z^+, \\
							\phi_0 ^T e^{i m_2 z}   						& \mathrm{if~~} z = z^-,
						\end{cases}
				\end{equation}
				where $\phi_0^I$, $\phi_0^R$, and $\phi_0^T$ are constants describing the amplitudes of the incoming, reflected, and transmitted waves, $m_1$ and $m_2$ are the vertical wavelengths corresponding to the media of buoyancy frequencies $N_1$ and $N_2$ defined by~\eqref{eq:IW04}, and $z^+$ and $z^-$ are two locations far from the interface respectively above and below it. The asymptotic expression $\tilde{\phi}$ can be written as boundary conditions for $\phi$
				\begin{subequations}
					\begin{align}
						\phi &= \tilde{\phi} \mathrm{~~~~~~~~~~~at~~} z = z^+, \\
						\phi' &= i m_2 \tilde{\phi} \mathrm{~~~~~~at~~} z = z^-.
					\end{align}
					\label{eq:IW09}
				\end{subequations}
				Thus, we obtain a boundary value problem that can be solved numerically, giving the $z$-dependent function $\phi$ and allowing for the computation the transmission and reflection coefficients.
	
% ..................................................................................... %
			\subsubsection{Weakly Viscous Correction}\label{viscous}
			
				As shown by \cite{boury2018}, viscous damping for radial modes has to be considered while using frequencies such as $\omega/N<0.5$. The weakly viscous correction can be described by assuming that the first order correction in $\varepsilon = \nu l^2 / \omega$ for a vertical dependence of the stream functions as $\phi(z) \propto e^{imz}$ takes the form \citep{boury2018}
				\begin{equation}
					m = m^{(0)} + i\varepsilon m^{(1)} + \mathcal{O}(\varepsilon^2),\label{eq:IW24}
				\end{equation}
				with $m^0$ being the inviscid vertical wave number defined by equation \eqref{eq:IW04}, and $m^1$ the correction term
				\begin{equation}
					i \varepsilon m^{(1)} = \mp \frac{i \varepsilon l}{2 (1 - \gamma^2) \alpha^3\sqrt{1-\alpha^2}},\label{eq:IW25}
				\end{equation}
				where $\alpha = \omega /N$ and $\gamma = f/\omega$. Hence, after propagating over a distance $z$, the weakly viscous streamfunction $\psi_\nu$ is written
			\begin{equation}
				\psi_\nu (z) = \psi (z) \exp (-\varepsilon m^{(1)} |z|).\label{eq:IW26}
			\end{equation}
			
				Using the above expressions, the transmission coefficient can be computed numerically and compared to the inviscid case. The problem is solved as a boundary value problem, using a similar system as proposed before~\eqref{eq:IW09}. Yet, as $\phi$ is now driven by a fourth order differential equation \citep{boury2018}, the boundary value problem requires four conditions to be closed
				\begin{subequations}
					\begin{align}
							\phi &= \tilde{\phi} &~&\mathrm{at~~} z = L, \\
							\phi'' &= - (m_1^{(0)} + \varepsilon m_1^{(1)}) ^2 \tilde{\phi} &~&\mathrm{at~~} z = L, \\
							\phi' &= i (m_2^{(0)} + \varepsilon m_2^{(1)}) \tilde{\phi} &~&\mathrm{at~~} z = z^-, \\
							\phi'' &= - (m_2^{(0)} + \varepsilon m_2^{(1)})^2 \tilde{\phi} &~&\mathrm{at~~} z = z^-.
					\end{align}
					\label{eq:IW10}
				\end{subequations}
			
% ------------------------------------------------------------------------------------- %
		\subsection{Transmission of a Confined Incident Wave}
			\subsubsection{Sharp Interface}
		
				In the previous section, semi-infinite domains were implicitly assumed. Such situations can be found in the atmosphere, as discontinuities in the buoyancy frequency profile are likely to be far from a source of internal waves. In the oceans, however, strong stratifications are often found close to the surface \citep{boury2018}, so the upper layer cannot be modeled by a region that extends vertically without boundaries.
				
				We therefore consider a sharp interface between $N_1$ and $N_2$ at $z=-L$, and an upper rigid boundary (the ocean surface) at $z=0$, as presented in figure~\ref{fig1}(b). Analytically, this problem can be solved as in the case of a semi-infinite domain, except that the boundary conditions are different because the incoming wave $\psi^I$ is not known. Instead, the total field at $z=-L$ is known and its vertical dependence satisfies: $\phi (z= -L) = \phi_0 = \phi_0 ^I e^{-i L m_1} + \phi_0 ^R  e^{i L m_1}$. The following transmission coefficients are derived using the same continuity properties at the interface as in the previous section
				\begin{align}
					T_{v_r} &= \left| \frac{v_r^T}{v_r^I} \right| = \left| \left[ \cos (L m_1) + i \left(\frac{m_1}{m_2}\right) \sin (L m_1) \right]^{-1} \right|, 
					\label{eq:IW11a} \\
					T_{v_z} &= \left| \frac{v_z^T}{v_z^I} \right| = \left| \left[ \cos (L m_1) + i \left(\frac{m_2}{m_1}\right) \sin (L m_1) \right]^{-1} \right|, 
					\label{eq:IW11b}\\
					\mathrm{and~~~}T_{j~} % &= \left[ \cos ^2 (L m_1) + i\left(\frac{m_1}{m_2} + \frac{m_2}{m_1}\right) \cos(L m_1) \sin (L m_1) - \sin ^2 (L m_1) \right]^{-1} \\
					&= \left| \left[ \cos (2 L m_1) + i \left(\frac{m_1^2 + m_2^2}{2 m_1 m_2}\right) \sin (2 L m_1) \right]^{-1} \right|,
					\label{eq:IW11c}\\
					\mathrm{with~~~} m_1 &= \pm l \left( \frac{N_1^2}{\omega^2}- 1\right)^{1/2} \mathrm{~~~~~~~and~~~~~~~} m_2 = \pm l \left( \frac{N_2^2}{\omega^2}- 1\right)^{1/2}.
					\label{eq:IW04bis}
				\end{align}
				where the $v_r^T$ and $v_z^T$ are the transmitted radial and vertical velocities respectively, and $v_r^I$ and $v_z^I$ are the incoming radial and vertical velocities for the total wavefield in the upper region (see figure~\ref{fig1}). All velocities are taken at the interface, which means in the limit $z\rightarrow -L$. In equation~\eqref{eq:IW04bis} we recall the result from equation~\eqref{eq:IW04} in the non-rotating case ($f=0$).
				
\begin{figure}
					\centering
					\epsfig{file=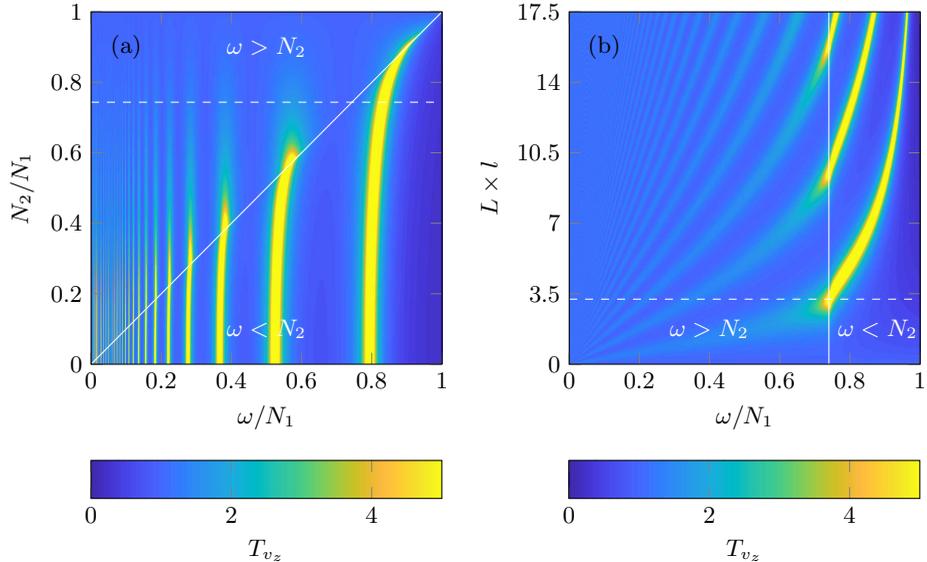} 
					\caption{Colormaps of the transmission coefficient $T_{v_z}$ as a function of (a) $\omega/N_1$ and $N_1/N_2$ at a given $L\times l=3.23$, and (b) $\omega/N_1$ and $L \times l$ at a given $N_2/N_1=0.74$. Dashed lines show the cuts at (a) $N_2/N_1=0.74$ and (b) $L\times l = 3.23$. The Coriolis frequency is set to be $f=0$.}
					\label{fig3}
				\end{figure}

				Expressions~\eqref{eq:IW11a},~\eqref{eq:IW11b}, and~\eqref{eq:IW11c}, invole different experimentally tunable variables: the buoyancy frequencies of the two layers $N_1$ and $N_2$, the frequency of the wave $\omega$, the depth $L$ of the interface, and the radial wave number $l$. These parameters can be reduced to three non-dimensionalised parameters: $L\times l$, $\omega/N_1$, and $N_2/N_1$. Figure~\ref{fig3} presents colormaps of the transmission coefficient for vertical velocity $T_{v_z}$, as a function of these different parameters. Figure~\ref{fig3}(a) shows the transmission coefficient as a function of $\omega/N_1$ and $N_2/N_1$ for a fixed value of $L\times l$ ($=3.23$). Transmission is generally higher for cases where waves can propagate in both regions ($\omega<N_2<N_1$) and drops off for evanescent frequencies in the lower region ($N_2 < \omega < N_1$). Maxima of transmission appear for well-defined values of $\omega/N_1$ at a given $N_2/N_1$. The curves defining maxima of transmission are continuous between the evanescent and the propagating regions. Figure~\ref{fig3}(b) shows the transmission coefficient as a function of $\omega/N_1$ and $L\times l$ for a fixed value of $N_2/N_1$ ($=0.74$). The cut-off between the propagating and evanescent regions is now observed at $\omega/N_1 = N_{1}/N_{2}=0.74$. The resonant peaks are expected to be less numerous as $N_2/N_1 \rightarrow 1$ or as $L\times l \rightarrow 0$. The study of the transmission coefficient in radial velocity $T_{v_r}$ shows similar behaviour.
		
% ..................................................................................... %
			\subsubsection{Smooth Interface}\label{smooth2}
			
				As in the freely incoming wave case, these results can be extended to the case of a smooth interface through the same method, simply by changing the boundary condition at the surface:
			\begin{equation}
				\tilde{\phi}(z) = 
					\begin{cases}
						\phi_0 	& \mathrm{if~~} z = L, \\
						\phi_0 ^T e^{i m_2 z} & \mathrm{if~~} z = z^-,
					\end{cases}
				\label{eq:IW12}
			\end{equation}
			where $z^-$ is a location far from the interface in the lower layer, leading to a new set of boundary conditions for the numerical computation.
			
% ..................................................................................... %
			\subsubsection{Weakly Viscous Correction}\label{viscous2}
			
				The transmission mechanism involved in the confined configuration is supposed to result from an infinite number of reflection in the upper layer. Such a condition, however, is not truly satisfied in the case of a viscous fluid, as the amplitude of the wave field decreases and vanishes after a finite number of reflections. Once again, the problem is solved numerically after finding out the correct boundary conditions, using the first order developpement of the vertical wave number $m$ \citep{boury2018}.
	
% ################################################################################ %
	\section{Experimental Apparatus}
	
\begin{figure}
			\centering
			\epsfig{file=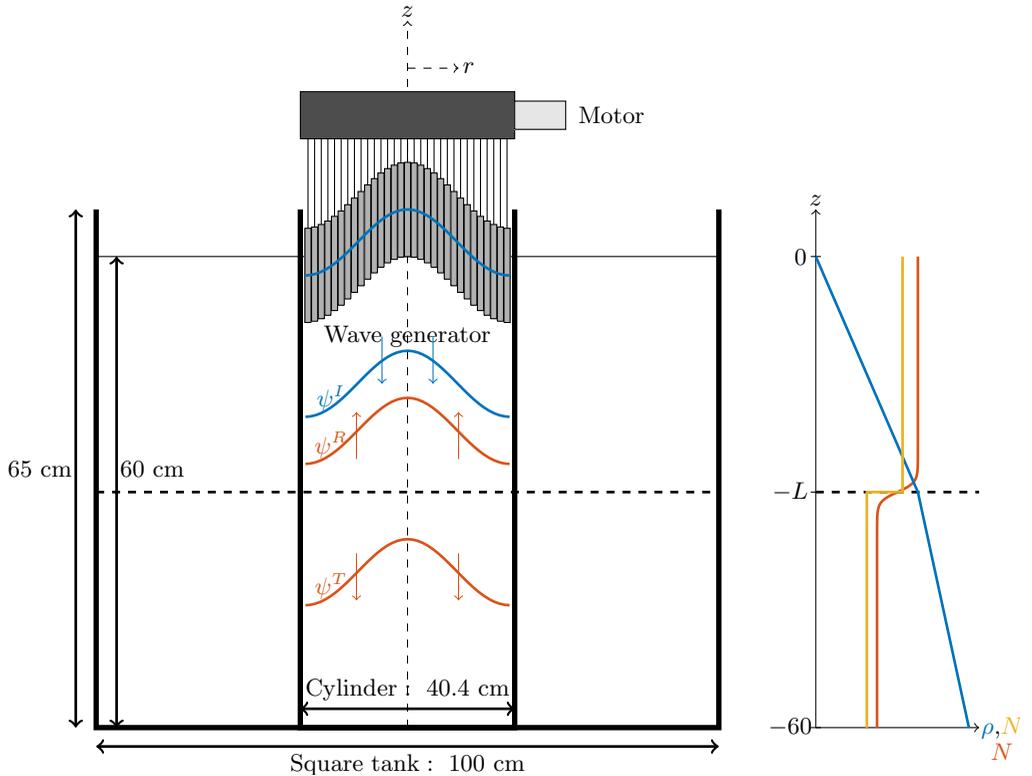} 
			\caption{Left: Schematic of the experimental apparatus. Right: Idealised density profile with corresponding sharp and smooth buoyancy frequency profiles.}
			\label{fig4}
		\end{figure}

		The experimental arrangement described in~\cite{boury2018} is used to investigate the transmission of internal waves across a buoyancy frequency interface. A schematic is presented in figure~\ref{fig4}. Experiments were conducted in a cylindrical acrylic tank of diameter $40.4\mathrm{~cm}$, inserted into a square acrylic tank of $100\mathrm{~cm}\times 100\mathrm{~cm}$ horizontal section and $65\mathrm{~cm}$ height. Both were filled with salt-stratified water with the same density profile to avoid optical deformations.
			
\begin{figure}
				\centering
				\epsfig{file=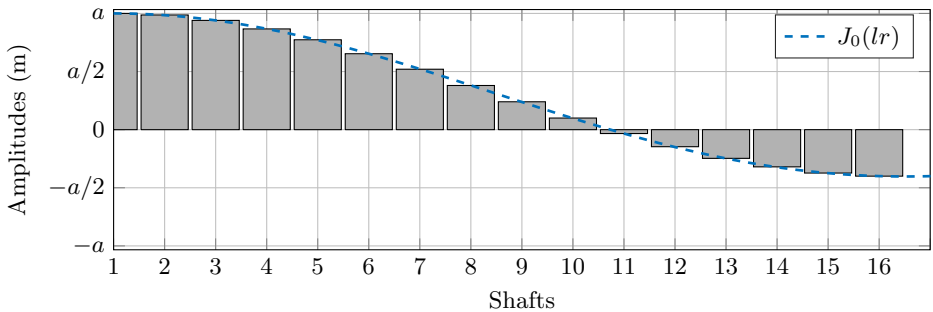} 
				\caption{Configuration of the generator ($l=19\mathrm{~m^{-1}}$). The amplitude $a$ is set to $2.5\mathrm{~mm}$.}
				\label{fig5}
			\end{figure}

			Internal waves are produced using Maurer~\textit{et al}'s axisymmetric wave generator~\citep{maurer2017}. This device, adapted from Gostiaux~\textit{et al}'s planar wave generator~\citep{gostiaux2006}, comprises $16$ concentric cylinders moving with customisable excentricities, enabling us to excite axisymmetric wave patterns. We set its configuration to be a mode $1$ profile able to produce pure cylindrical modes shaped as Bessel functions \citep{boury2018}. The profile of the mode $1$ configuration, with a radial wavelength of $l=19\mathrm{~m^{-1}}$, is presented in figure~\ref{fig5}. We use a generator amplitude $a=2.5\mathrm{~mm}$, which is low enough to limit non-linear effects.
			
			We adapted the double bucket method~\citep{fortuin1960, oster1963} to fill the tank with salt-stratified water. In order to obtain bi-linear stratifications, when required we stopped the filling, to change the filling tank density difference by removing salty water and adding fresh water without changing the density in the mixing tank. We then re-started the filling, which gave us a sharp but continuous buoyancy frequency interface at the desired depth. Density and buoyancy frequency profiles were measured using calibrated PME conductivity and temperature probes mounted on a motorized vertical axis. Buoyancy frequencies of each layer are estimated from the mean values of the $N$ profile, obtained after smoothing the density function $\rho(z)$ and taking its derivative. In our experiments, we decided to use a stronger stratification in the upper layer ($N_1>N_2$), as in the theory, because we wanted to look at transmission when the wave field in the lower layer was either propagating or evanescent.
			
			Visualisation of the velocity fields was performed using Particle Image Velocimetry (PIV). A horizontal laser plane, containing the generator diameter, was created using a $2\mathrm{~W}$ Ti:Sapphire laser (at $532\mathrm{~nm}$) and a cylindrical lens. Hollow glass spheres of $10\mathrm{~\mu m}$ diameter were added to the salt-stratified fluid for the purpose of visualisation. A camera recorded particle displacements at $1\mathrm{~Hz}$ in the laser plane. The CIVx algorithm was subsequently used to process the PIV raw images and extract the velocity fields~\citep{fincham2000}.
			
% ################################################################################ %
	\section{Results}
		\subsection{Transmission of a Free Incident Wave}
			
			We first quantified the transmission coefficient for the vertical velocity field in the case of a freely incoming wave. To conduct this experiment, a two-layer stratification is used and a range of forcing frequencies is explored. For each frequency, the transmission coefficient is extracted from the PIV data by looking at the vertical velocity below the interface.
			
\begin{figure}
				\centering
				\epsfig{file=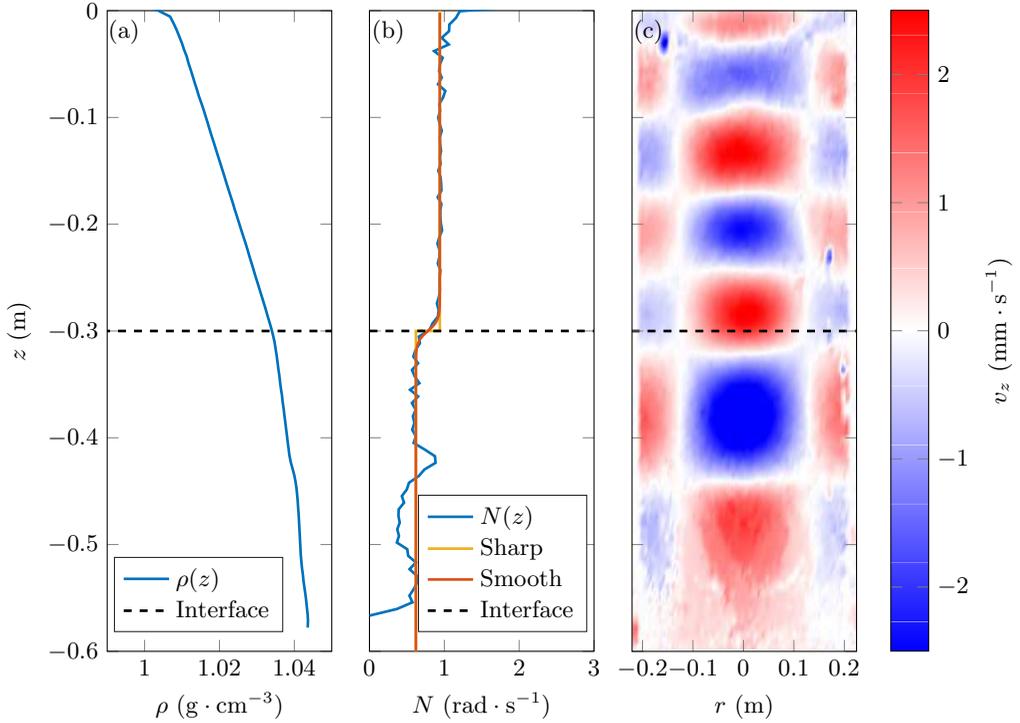} 
				\caption{From left to right: (a) density profile $\rho(z)$, (b) buoyancy frequency profile $N(z)$ and fits (with $N_1 = 0.94\pm 0.02 \mathrm{~rad\cdot s^{-1}}$, $N_2 = 0.62\pm 0.05 \mathrm{~rad\cdot s^{-1}}$, $L = 30\mathrm{~cm}$, width $\delta = 1\mathrm{~cm}$), and (c) example of vertical PIV cross-section of the vertical velocity field ($\omega/N_1 \simeq 0.4$). The straight line indicates the generator mean position, and the dashed line is the interface.}
				\label{fig6}
			\end{figure}

			Figure~\ref{fig6} presents the stratification used to conduct this study. The first two plots show the density profile $\rho (z)$ computed from the measurements of the C/T probe, and the buoyancy frequency profile $N(z)$ subsequently derived. The density curve shows two layers of linear stratification, one from $0$ to $30\mathrm{~cm}$ and one from $30$ to $60\mathrm{~cm}$. The buoyancy frequency profile confirms this trend, showing two constant values for the density gradient: $N_1 = 0.94\pm 0.02 \mathrm{~rad\cdot s^{-1}}$ and $N_2 = 0.62\pm 0.05 \mathrm{~rad\cdot s^{-1}}$. We indicate by a dashed line the interface between these two domains, and by a straight line the top of the tank. Two to three centimeters are missed in the measurements at the bottom of the tank, due to the configuration of the probe. On the buoyancy frequency curve, we fit the $N$ profile with our models for a sharp and a smooth interface (equation~\eqref{eq:IW07}) with the same distance $L=30\mathrm{~cm}$ from the generator, and a width $\delta = 1\mathrm{~cm}$. No error is given on these lengths because it does not impact the transmission coefficient error, as we will discuss later. A small bump in the stratification is present at $42\mathrm{~cm}$ below the generator: its width on the $N$ profile is exaggerated because of the averaging process used to compute the buoyancy frequency and, as can be seen in figure~\ref{fig6}(c), it does not affect the velocity field. However, to get the most accurate evaluation of $N_2$, we used the mean value and the standard deviation in the interval $[31;~41]\mathrm{~cm}$. The $N_1$ profile was estimated in the $[3;~28]\mathrm{~cm}$ interval.
			
			Figure~\ref{fig6}(c) presents a PIV snapshot of the vertical velocity field after the wave field crossed the buoyancy frequency interface. The mode$-1$ shape of the generator in a horizontal plane is visible at the top due to a parallax effect. Through the interface, we see that the radial wavelength, given by the horizontal nodes and antinodes of the field, is conserved. The vertical wave number, given by the vertical nodes and antinodes \citep{boury2018}, changes when the wave field crosses the interface, and is lower in the bottom layer than in the top one. Such a behaviour is expected, as the vertical wave number $m$ is fixed by the buoyancy frequency $N$ and in our experiment we set $N_2 < N_1$.
	
\begin{figure}
				\centering
				\epsfig{file=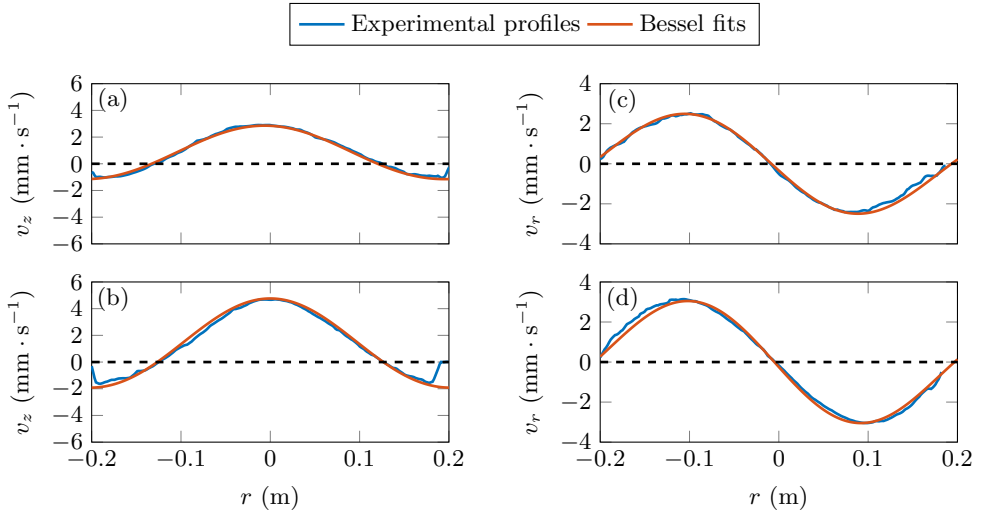} 
				\caption{Experimental profiles and fits by Bessel functions: (a) vertical velocity in the upper layer, at $z\simeq-15\mathrm{~cm}$; (b) vertical velocity in the lower layer, at $z\simeq-35\mathrm{~cm}$; (c) radial velocity in the upper layer, at $z\simeq-15\mathrm{~cm}$; (d) radial velocity in the lower layer, at $z\simeq-35\mathrm{~cm}$. Profiles are taken at random times.}
				\label{fig7}
			\end{figure}

	 	To ensure that the interface does not affect the modal shape of the field, we present in figure~\ref{fig7} horizontal profiles of the radial and vertical velocity fields in the upper layer and in the lower layer. Without any regard on the amplitude, as the profiles were taken at random times, we see that in both cases the fits with the expected Bessel functions respectively $J_0 (l r)$ and $J_1 (l r)$, with $l = 19\mathrm{~m^{-1}}$, agree well with the experimental profiles.
			
			The transmission properties are investigated using the stratification from figure~\ref{fig6}, by producing mode$-1$ wave fields first for increasing frequencies $\omega = 0.3$ to $0.98\mathrm{~rad\cdot s^{-1}}$ (labelled by $\omega_\nearrow$), then for decreasing frequencies $\omega = 0.96$ to $0.28\mathrm{~rad\cdot s^{-1}}$ (labelled by $\omega_\searrow$). As no direct measurement of the stratification can be performed inside the cylinder due to the lateral, top, and bottom boundaries, this protocol serves as a check that the stratification does not change while running the experiment. Because we only want to investigate the transmission of the wave field, short-time measurements are done with excitations of $100\mathrm{~s}$, at $20\mathrm{~min}$ intervals each, to ensure that the fluid is initially at rest for each measurement. We measure the amplitude in the lower region, after the wave has crossed the interface. To this extent, the selection procedure described in \cite{boury2018} is adapted. 
			
\begin{figure}
			\centering
			\epsfig{file=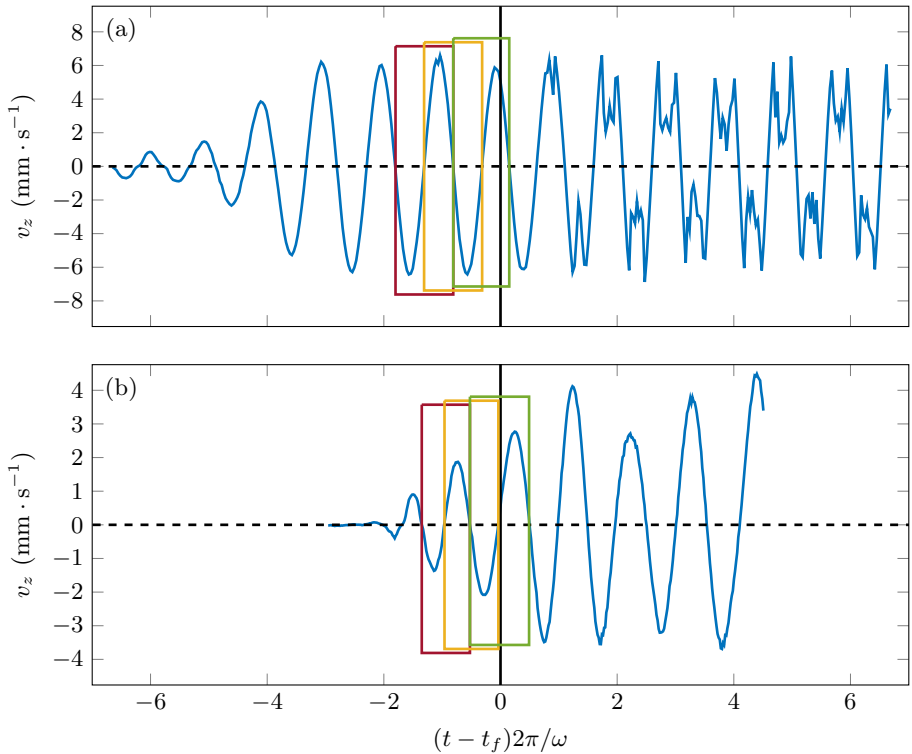} 
			\caption{Timeseries of the amplitude of the vertical velocity over $100\mathrm{~s}$, (a) at $\omega/N_1=0.89$ and (b) at $\omega/N_1=0.50$. The black solid line indicates the time $t_f$ used to compute the amplitude. Rectangles show the three periods used to estimate the wave amplitude from RMS.}
			\label{fig8}
		\end{figure}

		We consider the time series obtained by fitting horizontal cuts  of vertical velocity for Bessel functions of radial number $l=19\mathrm{~m^{-1}}$. These cuts are taken at a depth $z_m$, a few centimeters below the interface. In this time series, we pick three different time windows of one-period width, starting at a time when $v_z=0$, half-covering each other. The middle time window is chosen to be the last one before the wave reflected at the bottom returns at $z=z_m$. This returning time, $t_f$, is determined using the group velocity
			\begin{equation}
				\mathbf{v_g} = - \frac{m l^2 N^2}{\omega k^4} \mathbf{e_z},
				\label{eq:vg}
			\end{equation}
		computed in both layers. Figure~\ref{fig8} shows examples of (a) timeseries at $\omega/N_1=0.89$ and (b) timeseries at $\omega/N_1=0.50$ , with the three different time windows used to estimate the amplitude in both cases. The wave field in figure~\ref{fig8}(a) is evanescent in the lower region, so there is theoretically no reflected wave. In this case, the time $t_f$ is taken just before seeing non-linearities. In the measurement region, the temporal evolution of the amplitude is nearly sinusoidal. In figure~\ref{fig8}(b), the amplitude is still growing after the returning time. It can be due to the reflected wave, or to a non-fully established wave field.
			
			For each of the three time windows defined, the RMS value of the time signal during the chosen time windows is computed. The wave amplitude is then defined as the average of the three values obtained, multiplied by $\sqrt{2}$ since the signal is fairly sinusoidal. The standard deviation between these three values gives an estimate of the measurement error. The need of such a process is justified by the fact that for some frequencies, the wave field is not fully developped when reaching the interface (figure~\ref{fig8}). Given this situation, measuring the right amplitude is difficult even when the profiles are well fitted by the theoretical Bessel functions.		
		
		Direct comparison between the amplitude of the wave field above and under the interface is not possible, since it is difficult to measure the amplitude in the upper region, where incident and reflected wave get mixed. However, we have shown in~\cite{boury2018} that for frequencies above $\omega/N\simeq 0.3$, the experimental amplitude of an incident mode is well predicted by the generator amplitude $a\omega$, corrected with the theoretical viscous damping coefficient		
		\begin{equation}
			v_{z,\mathrm{inc}} = a\omega \times \exp(- m^{(1)}_1 L),
			\label{v_theo}
		\end{equation}
		where $m^{(1)}_1$ is defined in \ref{viscous}. Therefore, to estimate the transmission coefficient, we compare the measured transmitted amplitude $v_{z,\mathrm{mes}}$ to the theoretical incident amplitude $v_{z,\mathrm{inc}}$.		 
		\begin{equation}
			T_{v_z} = \frac{v_{z,\mathrm{mes}}}{v_{z,\mathrm{inc}}} = \frac{v_{z,\mathrm{mes}}}{a\omega \times \exp(- m^{(1)}_1 L)}.
		\end{equation}
			
\begin{figure}
				\centering
				\epsfig{file=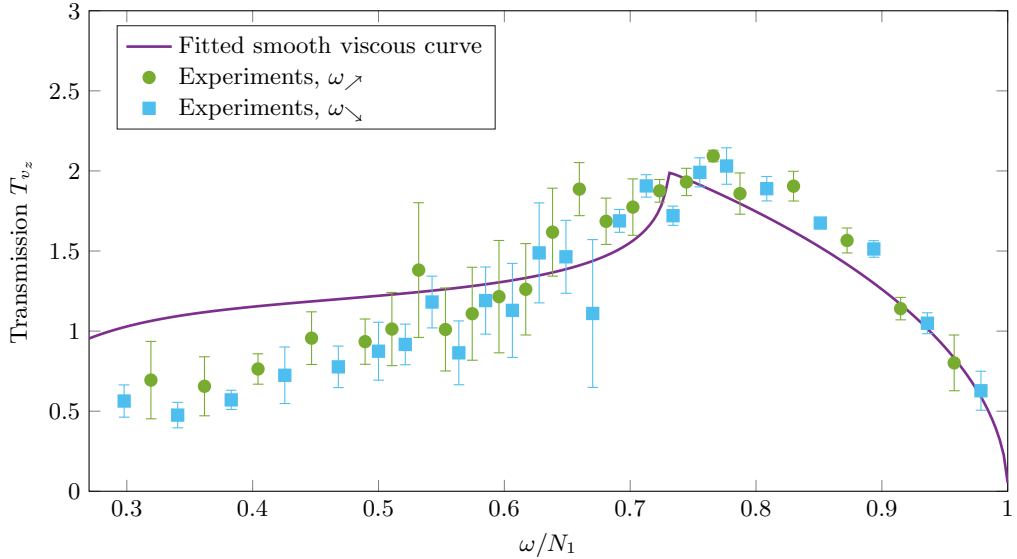} 
				\caption{Experimental transmission curve. We fit the experimental data with a theoretical curve computed for a smooth interface with viscosity (straight line).}
				\label{fig9}
			\end{figure}

			The experimental transmission data points are presented in figure~\ref{fig9}. To scale the frequencies as a non-dimensional parameter, we use the mean value of the buoyancy frequency $N_1$. We note that, as the two sets of measurements $\omega_\nearrow$ and $\omega_\searrow$ nicely follow the same trend, the influence of the forcing on the stratification can be considered as negligeable. In order to draw comparisons with theoretical predictions, we must first make some considerations on the control parameters. The shape of the theoretical curves is controlled by various parameters: the position of the interface $L$, its width $\delta$, and the buoyancy frequencies of the two layers $N_1$ and $N_2$. We studied the impact of these parameters by computing different curves for the smooth interface with a weakly viscous correction using the method explained in sections~\ref{smooth} and \ref{viscous}. We found out that the distance $L$ mainly controls the viscous damping observed at low frequencies, as the amplitude of the wave is greatly attenuated for $\omega/N_1 < 0.5$, as discussed in \cite{boury2018}. The width $\delta$ changes the height of the peak in the transmission curve, located at $\omega/N_1 = N_2/N_1$. However, as the interface is really sharp in our experiment ($\delta = 1\mathrm{~cm}$), its impact is negligeable. To investigate the influence of the boyancy frequencies, we looked at the ratio $N_2/N_1$. We found that the transmission curve is very sensitive to a slight change in the buoyancy frequencies, as it shifts the position of the transmission peak and defines the range of frequencies for which the wave is transmitted or fully reflected.
			
			As discussed at the beginning of this section, we measured $N_1 = 0.94\pm 0.02 \mathrm{~rad\cdot s^{-1}}$ and $N_2 = 0.62\pm 0.05 \mathrm{~rad\cdot s^{-1}}$. As a result, we have $N_2/N_1= 0.66\pm 0.06$. Du to the fairly large uncertainty on this value, however, we decided to compute numerically the transmission curves for a smooth interface with a weakly viscous correction, leaving the ratio $N_2/N_1$ as a free parameter that we adjust by ensuring that the position of the transmission peak in this curve corresponds to the data. This adjusted curve is shown in  figure~\ref{fig9}. The resulting fitted value of $N_2/N_1$ is 0.73, which is within the uncertainty of our experimental estimate for this ratio.

			We can now discuss how the experimental data points compare with the theory for a smooth interface and week viscous effects. The general trend is very similar, with an increased transmission as the frequency is increased, until one reaches a maximum of $T_{v_z}=2$ at a frequency close to the buoyancy frequency of the lower layer, as expected since the waves then become evanescent in this region. Then, one observes a sharp decrease of the transmission. Above $\omega/N_{1}\simeq 0.5$, the experimental data points show good agreement with the theoretical curve. In contrast, in the region $0.3<\omega/N_{1}<0.5$, the data points are below the theoretical prediction. We assume that this trend for low values of $\omega/N_1$ is due to the poor efficiency of the apparatus at these frequencies, because of boundary layer damping effects on the cylinder~\citep{beckebanze2018, boury2018}, which are not taken into account in our weakly viscous model.
		
% ------------------------------------------------------------------------------------- %
	\subsection{Transmission of a Confined Incident Wave}
	
\begin{figure}
				\centering
				\epsfig{file=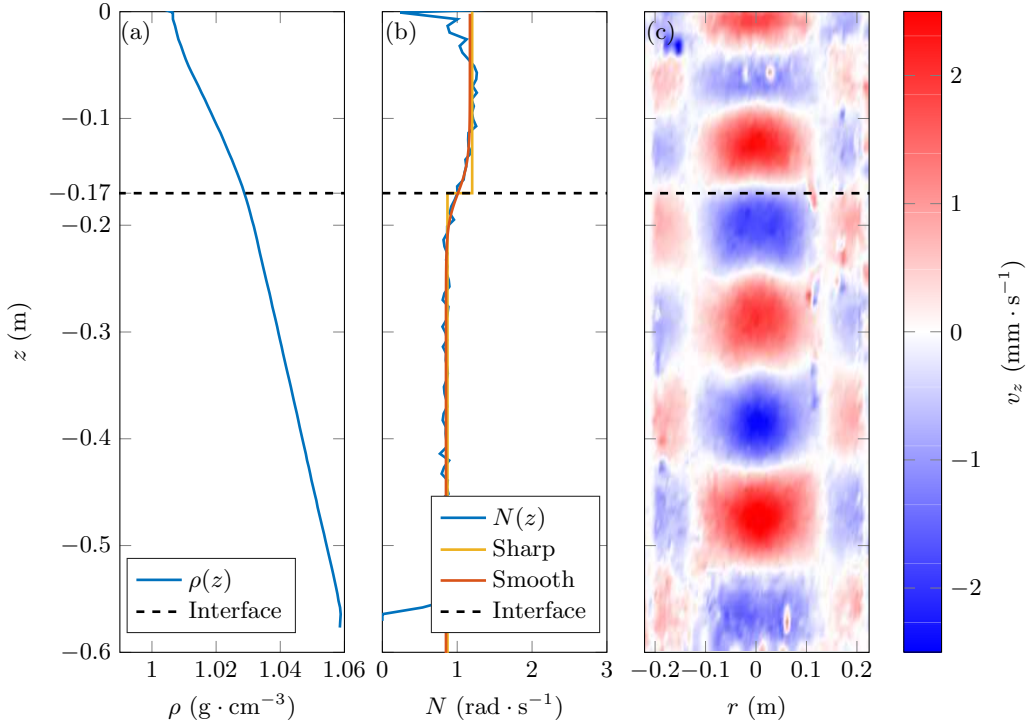} 
				\caption{From left to right: (a) density profile $\rho(z)$, (b) buoyancy frequency profile $N(z)$ and fits (with $N_1 = 1.17 \pm 0.07 \mathrm{~rad\cdot s^{-1}}$, $N_2 = 0.85 \pm 0.03 \mathrm{~rad\cdot s^{-1}}$, $L = 17\mathrm{~cm}$, width $\delta = 3\mathrm{~cm}$), and (c) example of vertical PIV cross-section of the vertical velocity field ($\omega/N_1 \simeq 0.36$). The straight line indicates the generator, and the dashed line the interface.}
				\label{fig10}
			\end{figure}

		We now focus on the confined wave configuration. Figure~\ref{fig10} shows the experimental stratification used in this study, with (a) the density and (b) the buoyancy frequency profiles, computed as in the previous section. We use a two-layer stratification: the top layer is from $0$ to $17\mathrm{~cm}$, with $N_1 = 1.17 \pm 0.07 \mathrm{~rad\cdot s^{-1}}$, and the bottom layer is from $17$ to $60\mathrm{~cm}$, with $N_2 = 0.85 \pm 0.03 \mathrm{~rad\cdot s^{-1}}$. A dashed line indicates the location of the interface, and a straight line shows the water surface. The buoyancy frequency profile is fitted with a sharp interface curve and a smooth interface curve, with the same distance $L=17\mathrm{~cm}$, and a width $\delta=3\mathrm{~cm}$ for the smooth interface. We see an important mixed layer at the surface, which is part of the upper layer and we make sure that $N_1$ is estimated outside this layer. We compute $N_1$ in the interval $[2;~15]\mathrm{~cm}$ and $N_2$ using the interval $[22;~53]\mathrm{~cm}$. Figure~\ref{fig10}(c) presents an example of a PIV field for the vertical velocity in this experiment. Between the generator, at the surface, and the interface, we can see that the wave field has a smaller vertical wavelength than below the interface, consistent with $N_2 < N_1$, as in the previous experiment.
		
		Mode$-1$ wave fields for $65$ values of frequency from $\omega = 0.2$ to $1.09 \mathrm{~rad\cdot s^{-1}}$ are produced. Given the previous results on transmission in a non-resonant case, we assume that the stratification does not change in our experiment and we only perform the measurements by increasing the frequency. We also use short-time excitations of $100\mathrm{~s}$ at $20\mathrm{~min}$ interval each. Amplitudes are measured below the interface in a similar selection process as before, but the position of the interface closer to the generator allows for several back and forth reflections in the upper layer. Measurements are compared to the theoretical generator amplitude~\eqref{v_theo} to compute the transmission coefficient.
		
\begin{figure}
			\centering
			\epsfig{file=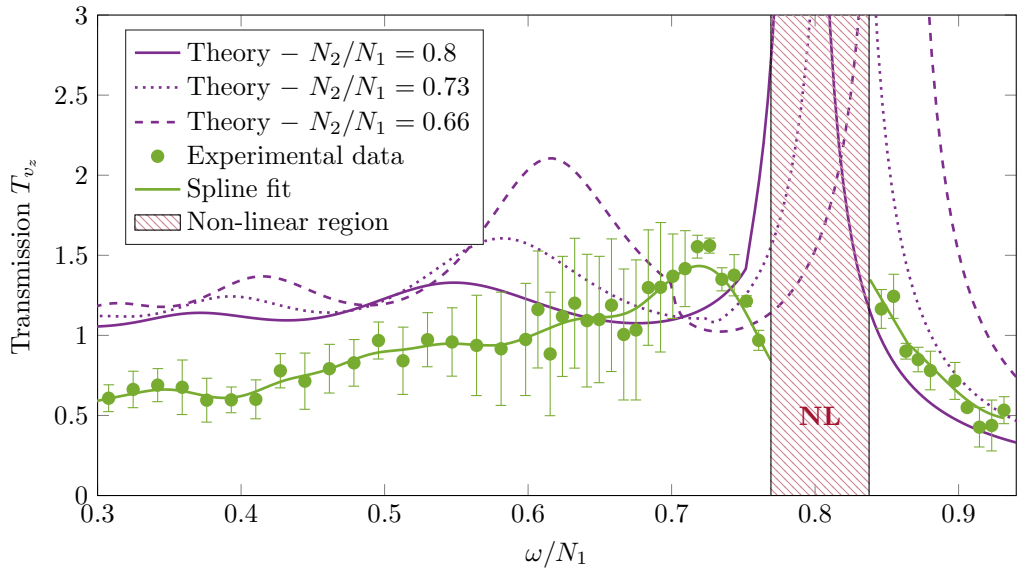} 
			\caption{Experimental transmission curve for a forced upper layer. Theoretical curves computed for the maximal, mean, and minimal values of $N_2/N_1$ are plotted. A spline fit (dashed green line) is added to guide the eye on the experimental curve.}
			\label{fig11}
		\end{figure}

		Results are presented in figure~\ref{fig11}. Measurements of the transmission coefficient are plotted as a function of $\omega/N_1$, using the mean value of $N_1$. As in the previous section, we must also investigate the influence of the different parameters $L$, $\delta$, $N_1$, and $N_2$, on the theoretical transmission curve for a confined incident wave, a smooth interface and weekly viscous effects obtained following the method described in section~\ref{smooth2} and \ref{viscous2}. The length $L$ contributes to the peak positions in the propagating region $\omega < N_2$.  The width of the interface $\delta$ has little impact because the interface is relatively sharp. In contrast, the buoyancy frequencies $N_1$ and $N_2$ mainly control the shape of the curve, as they change the position of the peaks, their amplitude, as well as the limit between the propagating and evanescent transmissions. Using the C/T probe, we measure $N_1 = 1.17 \pm 0.07 \mathrm{~rad\cdot s^{-1}}$ and $N_2 = 0.85 \pm 0.03 \mathrm{~rad\cdot s^{-1}}$, which gives us a ratio $N_2/N_1$ going from $0.8$ to $0.66$. As a result, several theoretical curves for a smooth interface with viscosity, numerically computed, are presented in figure~\ref{fig11}, one for the central value of $N_2/N_1$, namely 0.73, and two others for the extreme values of the ratio, 0.66 and 0.8. It shows that although the different theoretical curves always present the same trend, with various peaks, the position and height of the peaks can vary a lot by slightly varying the ratio $N_2/N_1$.
		
\begin{figure}
			\centering
			\epsfig{file=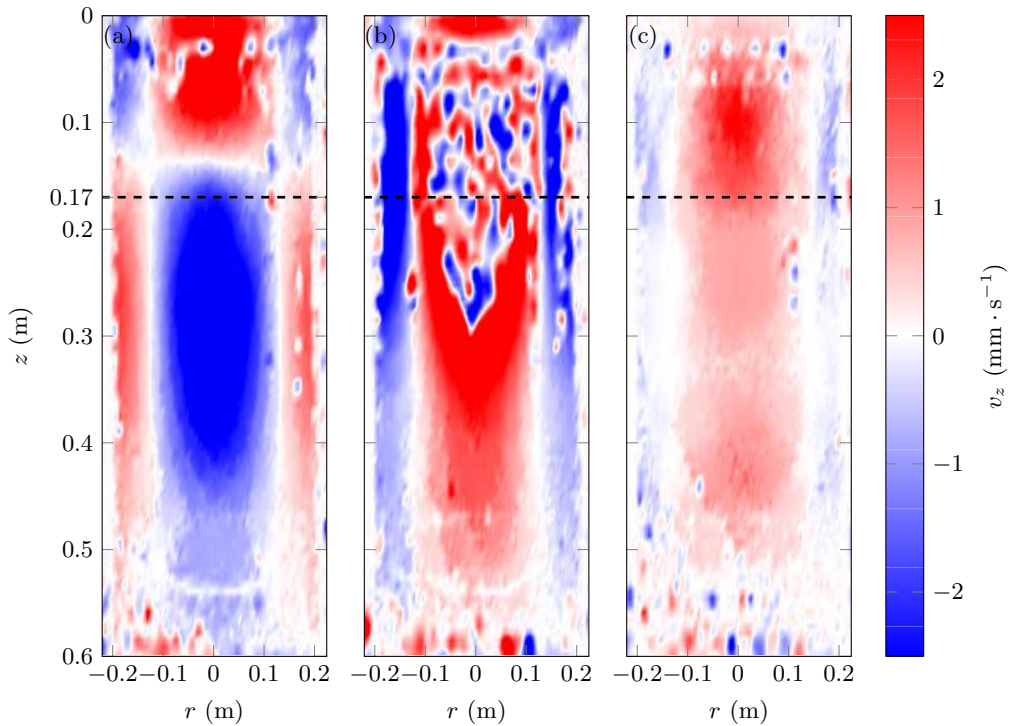} 
			\caption{Snapshots of the vertical velocity field in three different cases: (a) $\omega < N_2 < N_1$ and $\omega/N_1 = 0.7$; (b) $N_2 < \omega < N_1$ and $\omega/N_1 = 0.7$; and (c) $N_2 < \omega < N_1$ and $\omega/N_1 = 0.8$. All pictures are taken at $34$ periods of excitation.}
			\label{fig12}
		\end{figure}

		Experimental data display the same qualitative behaviour as theoretical predictions. For $\omega/N_1 \in [0.17;~0.74]$, the transmission coefficient increases non-monotonically with local extrema. As in the previous section, measurements at low frequencies show a smaller transmission than the expected value, also probably due to boundary layer damping effects~\citep{beckebanze2018, boury2018}. For $\omega/N_1 > 0.74$ the transmission coefficient is globally decreasing as the waves become evanescent in the lower region. For a given range of frequencies $\omega/N_1 \in [0.77;~0.84]$, however, non-linear effects are triggered and no amplitude can be measured as the wave field is no longer described by our model. This area is indicated by a shaded area in figure~\ref{fig11} and corresponds to the expected resonance peak of the theoretical curves. Snapshot examples of the vertical velocity field are given in figure~\ref{fig12} for three different situations after $34$ periods of excitation: figure~\ref{fig12}(a) shows propagating waves in both layers at $\omega/N_1 = 0.7$, while in figure~\ref{fig12}(b) and~\ref{fig12}(c) waves are evanescent in the lower layer at $\omega/N_1 = 0.8$ and $\omega/N_1 = 0.9$, respectively. In contrast to the images presented in figures~\ref{fig12}(a) and~\ref{fig12}(c), the image presented in figure~\ref{fig12}(b) shows small-scale disturbances of the wave field resulting from non-linear effects due to resonance in the upper layer, representing the typical behaviour observed within the hatched region identified in figure~\ref{fig11}.
		
% ################################################################################ %
	\section{Application to ocean: an example}
	
\begin{figure}
			\centering
			\epsfig{file=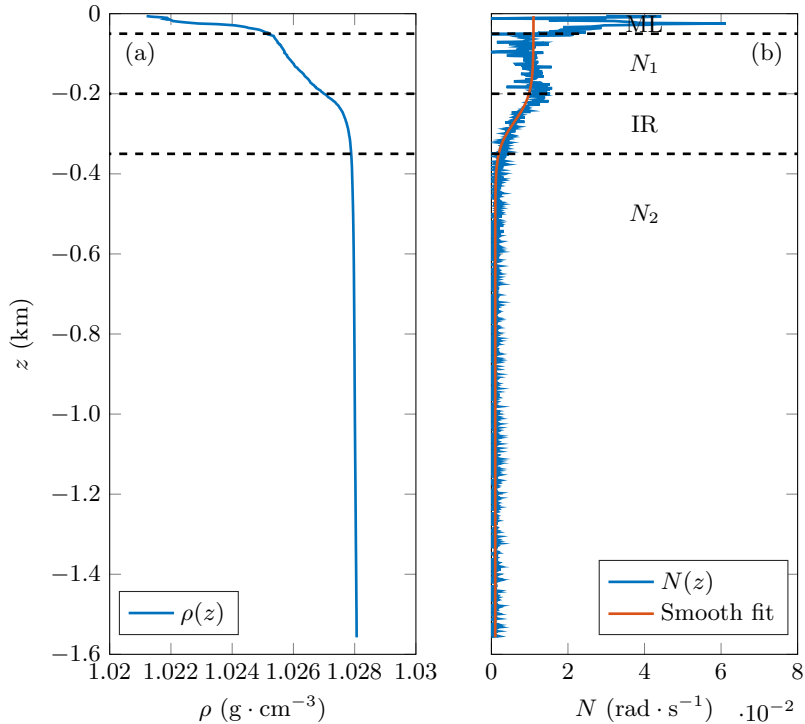} 
			\caption{Example of (a) density profile and (b) buoyancy profile, from in-situ measurements. Different layers can be identified between dashed lines, from top to bottom: Melting Layer (ML), $N_1$ linear layer ($N_1$), Interface Region (IR), and $N_2$ linear layer ($N_2$).}
			\label{fig13}
		\end{figure}

		An interesting application of our confined layer studies is to relate our findings to a scenario that has a strong upper ocean stratification overlying a weaker deep ocean stratification. Figure~\ref{fig13} presents an example of density and buoyancy frequency profiles taken at $ 159 ^\circ 57.111' \mathrm{~W}$, $ 73^\circ 32.439' \mathrm{~N}$ during the Stratified Ocean Dynamics of the Arctic (SODA) research cruise, on the R/V Sikuliaq, in September $2018$. A very simple model can be used to describe this stratification at first order: a Melting Layer (ML) at the surface issued from ice melting with a strong density gradient, and a bi-linear stratification with two buoyancy frequencies $N_1$ and $N_2$ below. Using equation~\eqref{eq:IW07}, such a profile can be fitted with $N_1 \simeq 0.011 \mathrm{~rad\cdot s^{-1}}$ in the upper layer, $N_2 \simeq 0.001 \mathrm{~rad\cdot s^{-1}}$ in the lower layer, $L \simeq 250\mathrm{~m}$ from the Melting Layer (ML) to the center of the Interface Region (IR), and $\delta \simeq 150\mathrm{~m}$ for the width of the Interface (IR).
		
\begin{figure}
			\centering
			\epsfig{file=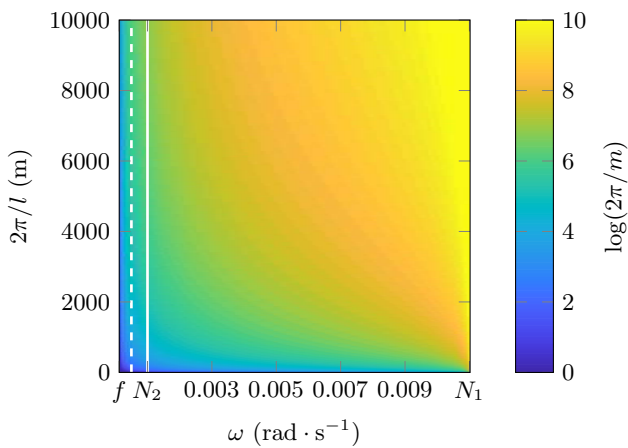} 
			\caption{Logarithmic colormap of the vertical wavelength in the upper layer ($N_1 = 0.011\mathrm{~rad\cdot s^{-1}}$) as a function of the frequency $\omega$ (from $f$ to $N_1$) and the radial wavelength $2\pi/l$. The white solid line represents $\omega=N_2$, and the white dashed line represents $\omega=N_2/2$.}
			\label{fig14}
		\end{figure}

		As shown by our study, resonance effects may appear and produce high transmission coefficient value for confined internal waves. These effects, however, arise only when the vertical wavelength $2\pi / m_1$ is larger than the interface width \citep{mathur2009}. Upon the contrary, the interface acts as a smooth density change for the wave and internal waves pass through relatively unaffected. Re-writing equation~\eqref{eq:IW04} in terms of the radial wavelength $2\pi / l$ as
		\begin{equation}
			\frac{2 \pi}{m_1} = \frac{2 \pi}{l} \left(\frac{\omega^2 - f^2}{N_1^2 - \omega^2} \right)^{1/2},
		\end{equation}
		we explore the range of parameters $(\omega;~2\pi/l)$ by plotting the quantity $\log (2\pi/m_1)$ for propagating waves in the upper layer, which means $f < \omega < N_1$, in figure~\ref{fig14}. The Coriolis frequency is set to be $f = 1.24\cdot 10^{-4}\mathrm{~rad\cdot s^{-1}}$, a typical value in this region~\citep{cole2013}.
		
		Near-inertial wave generation, with frequencies close to $f$ are found to be easily excited in the Arctic Ocean~\citep{cole2013}, with a small vertical wavelength as illustrated in figure~\ref{fig14}. In this configuration, the factor $\omega^2 - f^2$ is of the order of $f^2$ ($10^{-8}\mathrm{~rad^2\cdot s^{-2}}$). Since $N_1 \gg \omega$, the factor $N_1^2 - \omega^2$ is of the order of $N_1^2$ ($10^{-4}\mathrm{~rad^2\cdot s^{-2}}$), yielding $l / m_1 \simeq 10^{-2}$. Hence, to obtain vertical wavelengths of the order of $100\mathrm{~m}$, the radial wavelength has to be of the order of $10\mathrm{~km}$, which can be produced by storms of large extent. Higher frequency waves can also be produced~\citep{bell1978, chini2005, polton2008}, with a frequency of the same order of magnitude than the buoyancy frequency $N_2$. Hence, the term $\omega^2 - f^2$ is of the order of $N_2^2$ ($10^{-6}\mathrm{~rad^2\cdot s^{-2}}$), yielding $l/m_1 \simeq 10^{-1}$. As a result, resonant waves with vertical wavelength of the order of $100\mathrm{~m}$ can have small radial wavelength (of the order of $1\mathrm{~km}$).
		
\begin{figure}
			\centering
			\epsfig{file=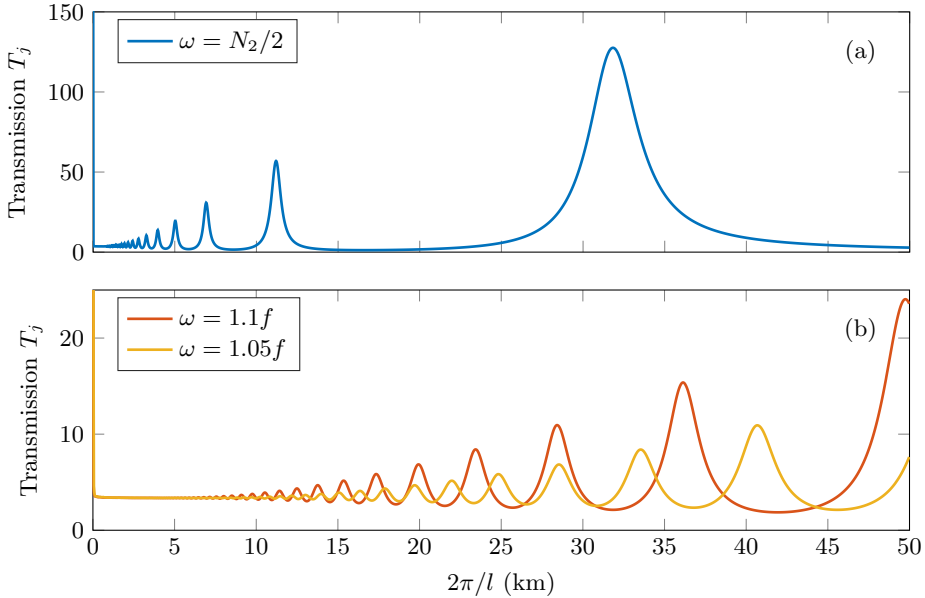} 
			\caption{Numerical computation of the transmission coefficient as a function of the radial wavelength, for three different frequencies $1.05 f$, $1.1 f$, and $N_2/2$, after normalisation by the incoming energy at the surface.}
			\label{fig15}
		\end{figure}

		Following the approach of~\cite{ghaemsaidi2016}, we present in figure~\ref{fig15} numerical computation of the transmission coefficient for the Arctic stratification (figure~\ref{fig13}) for three different frequencies, two of them being near-inertial at $\omega=1.05f$ and $\omega=1.1 f$, and the third one corresponding to a higher frequency with $\omega = N_2/2$. A resonance effect, with sequential peaks of high transmission rate of internal waves, occurs for $\omega=N_2/2$, for both small or large radial wavelengths. For near-inertial waves, however, the transmission coefficient is almost constant until the vertical wavelength is larger than $10\mathrm{~km}$, meaning that this enhancement is only relevant for large wavelength storms. In both scenarios the enhancement can be several times, even over an order of magnitude. 

% ################################################################################ %
	\section{Conclusions and Discussion}
	
		We have presented an experimental study of the propagation of axisymmetric internal wave modes across a buoyancy interface. To support our laboratory experiments, we develop a theoretical framework for freely incoming wave fields and for confined wave fields, as they represent qualitatively two different scenarios that have relevance to the atmosphere and ocean, respectively.
	
		For a free incident wave, we show that the maximum of transmission occurs when the frequency of the wave is equal to the buoyancy frequency of the lower layer. For a confined incident wave, we identify the existence of a resonance effect in the upper layer, leading to larger transmission coefficients and potentially non-linear effects. For both scenarios, theory and experiments show excellent qualitative, and good quantitative, agreement.
		
		This kind of study can help to shed light on in-situ measurements of internal waves signals, for example near-inertial waves in strongly stratified regions such as the Arctic Ocean. Different scenarios can be investigated, such as high frequency internal waves generated by storms at the ocean surface and travelling downwards, or near-inertial waves produced by tides and topography in the deep ocean and travelling upwards. In particular, this latter type of waves can therefore be enhanced through transmission processes and lead to strong signals though the generation process might be of very low amplitude.
		
		In the same way small-extent staircase stratifications can create strong resonant penetration of internal waves~\citep{ghaemsaidi2016}, we show that larger non-linear stratifications can potentially give rise to high transmission of internal waves to the deep ocean. Nevertheless, for small scale waves of wavelength below $10\mathrm{~km}$, these effects seem to be limited to high frequencies, while no effect is observed for near-inertial waves. For the latter type of waves, enhanced transmission can however also be observed in the case of very large size disturbances of wavelength above $20\mathrm{~km}$. We notice that the process involves higher amplification factors as frequency increases, while becoming more selective in terms of wavelength.
		
		Such a study could be extended to more complex cases with various layers of different buoyancy frequencies, as long as the thickness of the interfaces remains small compared to the vertical wavelength~\citep{nault2007, ghaemsaidi2016}. For a given wave, generated by a storm and observed near the ocean surface, it could help predict the amount of energy that will be carried out towards the deep ocean, as well as the ranges of wavelengths more likely to be transmitted. Such a selection process would lead to a change in the wave features as it travels downwards. Enhanced amplitudes generated by such resonant behaviours can also trigger non-linear effets~\citep{boury2018} producing smaller scale waves that could lead to mixing events, changing the stratification, with subsequent feed-back effets on the wave propagation.

\bigskip
\textbf{Acknowledgments}

This work has been partially supported by the ANR through grant ANR-17-CE30-0003 (DisET) and by ONR Physical Oceanography Grant N000141612450. S.B. wants to thank Labex iMust for supporting his research. T.P. thanks ENS de Lyon for travel support funding. The authors also thank R. Supekar for helpful discussions and inputs.

% ################################################################################ %

%
%\bibliographystyle{jfm}
%% Note the spaces between the initials
%\bibliography{P2-biblio}

\end{document}